\documentclass[10pt,preprint2]{aastex}
\usepackage{bm}
\usepackage{textcomp}

\shorttitle{$F$-state interference in a
non-magnetic medium}
\shortauthors{Smitha et al.}

\begin{document}
\title{{\bf Polarized line transfer with $F$-state interference in a
non-magnetic medium\\
{\it{Partial frequency redistribution  effects
in the collisionless regime}}}}

\author{H. N. Smitha$^{1}$, K. Sowmya$^{1}$, K. N. Nagendra$^{1}$,
M. Sampoorna$^{1}$ and J. O. Stenflo$^{2,3}$} 
\affil{$^1$Indian Institute of Astrophysics, Koramangala,
Bangalore, India}
\affil{$^2$Institute of Astronomy, ETH Zurich,
CH-8093 \  Zurich, Switzerland }
\affil{$^3$Istituto Ricerche Solari Locarno, Via Patocchi,
6605 Locarno-Monti, Switzerland}

\email{smithahn@iiap.res.in; ksowmya@iiap.res.in; knn@iiap.res.in;
sampoorna@iiap.res.in; stenflo@astro.phys.ethz.ch}
\begin{abstract}
Quantum interference phenomena manifests itself
in several ways in the polarized solar spectrum formed due to 
coherent scattering processes. One such effect
arises due to interference between the fine structure $(J)$ states
giving rise to multiplets. Another effect is that which arises due to
interference between the hyperfine structure $(F)$ states. 
We extend the redistribution matrix derived for the
$J$-state interference to the case of $F$-state interference.
We then incorporate it into the polarized radiative transfer
equation and solve it for isothermal constant property slab
atmospheres.
The relevant transfer equation is solved using
a polarized approximate lambda iteration (PALI) technique based on
operator perturbation. An alternative method derived from the
Neumann series expansion is also proposed and is found to be
relatively more efficient than the PALI method. The effects of PRD and the $F$-state
interference on the shapes of the linearly polarized Stokes 
profiles are discussed.
The emergent Stokes profiles are computed for 
hypothetical line transitions arising due to hyperfine 
structure splitting (HFS) of the upper 
$J=3/2$ and lower $J=1/2$ levels of a two-level atom model
with nuclear spin $I_s=3/2$. We confine our attention to the 
non-magnetic scattering in the collisionless regime.

\end{abstract}
\maketitle
\section{Introduction}
\label{intro}
\begin{figure}
\centering
\includegraphics[width=4.0cm,height=3.0cm]{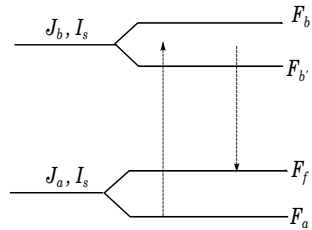}
\caption{\footnotesize Level diagram representing the 
HFS in a two-level atom model.}
\label{level-diag}
\end{figure}

The linearly polarized solar spectrum is produced by coherent
scattering processes taking place in the solar atmosphere. This
so called `second solar spectrum' is highly structured and
reveals various physical processes responsible to generate
the polarized signals in the spectrum. Quantum interference
is one such physical process whose importance has been
highlighted in the second solar spectrum studies
\citep[see][]{s80,jos94,s97}.
The coherent superposition of the fine structure states leads to 
the $J$-state interference, whereas the $F$-state interference
arises due to superposition of the hyperfine structure states 
(see Figure~{\ref{level-diag}}).
The $J$-state interference theory for the case of frequency
coherent scattering was developed by \citet[][]{s80,jos94,s97}. This theory was
extended to include partial frequency redistribution (PRD) in
line scattering, by \citet[][hereafter called P1]{smi11a}.
The $J$-state PRD matrix derived in P1 is used in the polarized line transfer
equation in \citet[][herafter P2]{smi11b}.
An alternative scattering
theory of $J$-state interference based on metalevel approach
was developed by \citet[][]{landietal97}, which also includes the
$F$-state interference effects. All the papers mentioned so far
are applicable to the case of colisionless regime.

Second solar spectrum contains several lines which have signatures
of $F$-state interference. Examples of these lines
are Na {\sc i} D$_2$ at 5890\,\AA, Ba {\sc ii} D$_2$ at 4554\,\AA , 
Mn {\sc i} 8741\ \AA, Sc {\sc ii} 4247\,\AA\ etc.
In this paper we are concerned with the line formation studies 
involving $F$-state interference process and PRD.
The $F$-state redistribution matrix derived in this 
paper can be used for modeling the non-magnetic 
quiet region observations of hyperfine structure splitting (HFS)
in the lines mentioned above.

The $F$-state interference theory applicable to the frequency 
coherent scattering was developed by \citet[][]{s97}. This 
theory, along with PRD, was applied by \citet[][]{fluri03} 
and \citet[][]{holz05} in the polarized line transfer
computations.
In \citet[][]{ll04} the theory of $F$-state interference was
developed under the approximation of complete frequency
redistribution (CRD).
The theory of $F$-state interference in a magnetic field
for multi-term atoms in the collisionless regime
and under the approximation of CRD is presented in 
\citet[][]{manso05}.

In the present paper we extend the $J$-state
interference theory presented in P1 to the case of $F$-state
interference. 
The $F$-state redistribution matrix is derived here for the non-magnetic case
and in the collisionless regime. The reason for considering the 
non-magnetic case in this paper, is that the formulation of P1 was confined
to the linear Zeeman regime of field strengths (the spacing between the Zeeman 
$m$-states being smaller than the spacing between the fine structure states).
In the present context, the hyperfine splitting becomes comparable to the 
Zeeman splitting even for weak magnetic fields, and we quickly enter the 
Paschen-Back regime of field strengths (level crossing of the $m$-states
belonging to different $F$-states), in which the formulation 
presented in P1 is not valid.
Since the Paschen-Back effect is outside the scope of the present paper, 
our treatment here is limited to the non-magnetic case, but the extension 
to the Paschen-Back regime is planned to be pursued in future work. We further
 assume that the lower level is unpolarized and infinitely sharp. While this 
assumption is made for the sake of mathematical simplicity, it is physically 
justified for the long-lived ground states, which are correspondingly more 
vulnerable to collisional depolarization.

Following P2, this PRD matrix is incorporated 
into the polarized line transfer equation, and solved using an 
operator perturbation method.
We also propose a new method to solve
the $F$-state interference problem. It is called the scattering
expansion method (SEM) and is described in \citet[][]{hfetal09,sam11}.
Recently it has been applied to a variety of problems 
\citep[see][]{sow12,sup12}.
We compare the operator perturbation method and the SEM by
applying them to the problem at hand.

In Section~{\ref{theory}} we derive the PRD matrix for $F$-state 
interference and incorporate it into the line transfer equation. 
In Section~{\ref{num-method}} we describe the numerical methods 
used to solve the transfer equation. Results are presented in 
Section~{\ref{results}}. Section~{\ref{conclusions}}
is devoted to the concluding remarks. 
\section{Basic equations}
\label{theory}

\subsection{The redistribution matrix}
\label{redmatrix}
In this section we present the redistribution matrix for the
$F$-state interference, derived starting from the
Kramers-Heisenberg formula. We restrict our attention to the
case of non-magnetic collisionless regime.

The redistribution matrix for the $F$-state interference can
be derived through a straight forward replacement of quantum
numbers, in the $J$-state interference redistribution matrix
derived in P1. The replacements are as follows
\citep[see][]{s97,ll04}:
\begin{equation}
L\rightarrow J;\ \ \ J\rightarrow F;\ \ \ S\rightarrow I_s,
\label{quantum} 
\end{equation}
where $L$, $J$ and $S$ represent the orbital, electronic, and
spin quantum numbers of a given state. $F$ is the total
angular momentum and $I_s$ the nuclear spin of the atom under
consideration. The expression for the $F$-state interference
redistribution matrix expressed in terms of irreducible
spherical tensors can be written as
\begin{eqnarray}
&&{\bf R}_{ij}^{\rm II}(x,{\bm n};x^\prime,
{\bm n}^\prime) = 
\frac{3(2J_b+1)}{2I_s+1}\nonumber \\&&
\times \sum_{KF_aF_fF_bF_{b^\prime}} (-1)^{F_f-F_a}
\cos\beta_{F_{b^\prime}F_b} 
{\rm e}^{{\rm i}\beta_{F_{b^\prime}F_b}}\nonumber \\ &&
\times \Big[(h^{\rm II}_{F_b,F_{b^\prime}})_{F_aF_f}
+{\rm i}(f^{\rm II}_{F_b,F_{b^\prime}})_{F_aF_f}\Big]
\nonumber \\ && \times
 (2F_a+1)(2F_f+1)(2F_b+1)(2F_{b^\prime}+1) \nonumber
\\&& \times \left\lbrace 
\begin{array}{ccc}
J_a & J_b & 1\\
F_b & F_f & I_s \\
\end{array}
\right\rbrace
\left\lbrace 
\begin{array}{ccc}
J_a & J_b & 1\\
F_b & F_a & I_s \\
\end{array}
\right\rbrace \nonumber \\&& 
\times \left\lbrace 
\begin{array}{ccc}
J_a & J_b & 1\\
F_{b^\prime} & F_f & I_s \\
\end{array}
\right\rbrace 
\left\lbrace 
\begin{array}{ccc}
J_a & J_b & 1\\
F_{b^\prime} & F_a & I_s \\
\end{array}
\right\rbrace \nonumber \\ && \times
\left\{
\begin{array}{ccc}
1 & 1 & K \\
F_{b^\prime} & F_b & F_a \\
\end{array}
\right\}
\left\{
\begin{array}{ccc}
1 & 1 & K \\
F_{b^\prime} & F_b & F_f \\
\end{array}
\right\} \nonumber \\&&
\times {\mathcal T}^K_0(i,{\bm n})
{\mathcal T}^{K}_{0}(j,{\bm n}^\prime).
\label{red_mat}
\end{eqnarray}
In the above expression the angle $\beta_{F_{b^\prime}F_b}$ is
defined as
\begin{equation}
 \tan\beta_{F_{b^\prime}F_b} = \frac{\omega_{F_{b^\prime}F_b}}{\gamma},
\label{han_ang} 
\end{equation}
where $\hbar\omega_{F_{b^\prime}F_b}$ represent the energy differences
between the $F_{b^\prime}$ and $F_b$ states in the absence of a
magnetic field. $\gamma$ is the damping parameter of the upper state.
The lower levels are assumed to be infinitely sharp and unpolarized.
The $h$ and $f$ functions are the auxiliary quantities defined in the
same way as Equations~(14) - (15) of P1, but with the replacements given in
Equation~(\ref{quantum}). ${\mathcal T}^K_Q$ are the irreducible tensors
for polarimetry introduced by \citet[][]{lan84}. For the non-magnetic
case presented in this paper $Q=0$. The indices $i$ and $j$
refer to the Stokes parameters ($i,j=0,1,2,3$) with $K=0,1,2$ and
$-K\leq Q \leq +K$. The directions of the 
incoming and scattered rays are given by  ${\bm n}^\prime$
 and $\bm n$ respectively.
 ${\bm n}=(\theta,\varphi)$ with
$\theta$ being the colatitude and $\varphi$ being the azimuth of
 the outgoing ray.
$x^\prime$  and $x$ are the incoming and scattered frequencies in 
Doppler width units.

\subsection{The polarized line transfer equation}
The one dimensional radiative transfer equation for solving the 
line formation problems with PRD and $F$-state interference
in scattering in the absence of a magnetic field is given by
\begin{equation}
 \mu\frac{\partial{\bm I}(\tau,x,\mu)}{\partial \tau}= (\phi_{\rm HFS}(x) + 
{r})[{\bm I}(\tau, x, \mu)-{\bm S}(\tau, x,\mu)],
\label{rte-1}
\end{equation}
where ${\bm I}=(I,Q)^{\rm T}$ is the Stokes vector and 
${\bm S}=(S_I,S_Q)^{\rm T}$ is the Stokes source vector.
Equation~(\ref{rte-1}) is valid for the case
of a two-level atom with an infinitely sharp 
and unpolarized ground level.
$\mu={\rm cos}\,\theta$ represents the line of sight.
$r$ is the ratio of continuum to the frequency-integrated line absorption
coefficient.
The positive Stokes $Q$ represents electric
vector vibrations perpendicular to the solar limb.
$\tau$ is the line
optical depth defined by $d\tau=-k_Ldz$, where $k_L$ is the 
frequency-integrated 
line absorption coefficient defined for a two-level atom
with HFS. If $\eta_L$ is the
 line absorption
coefficient then for the standard two-level atom without HFS,
 $\eta_L=k_L\phi(x)$ where $\phi(x)$ is the 
Voigt profile function. In the presence of HFS,
$\eta_L$ is given by (see Equation~(7) of P2)
\begin{eqnarray}
&&\eta_{L}(\nu)=\frac{k_L}{(2I_s+1)}\sum_{F_aF_b}
(2F_a+1)(2F_b+1)\nonumber \\ && \times
\left\lbrace
\begin{array}{ccc}
J_b & J_a & 1\\
F_a & F_b & I_s \\
\end{array}
\right\rbrace ^{2}
\phi(\nu_{F_bF_a}-\nu),
\label{comb-prof-func}
\end{eqnarray}
where
\begin{equation}
k_L=\frac{h\nu_{J_bJ_a}}{4\pi}N(J_a)B(J_a\to J_b),
\end{equation}
is the frequency-integrated absorption coefficient 
for all the $F$-states.
Thus $\phi_{\rm HFS}(x)$ is the weighted sum 
of the individual Voigt profiles $\phi(\nu_{F_bF_a}-\nu)$
representing
each of the $F_a \to F_b$ absorption.

For the particular case of $J_a=1/2 \to J_b=3/2 \to J_f=1/2$
transition with $I_s=3/2$, $\phi_{\rm HFS}(x)$ takes the form
\begin{eqnarray}
&&\phi_{\rm HFS}(x)= \bigg[\frac{2}{32}\phi(\nu_{0\,1}-\nu)+ 
\frac{5}{32}\phi(\nu_{1\,1}-\nu)\nonumber \\ && +
\frac{5}{32}\phi(\nu_{2\,1}-\nu) + 
\frac{1}{32}\phi(\nu_{1\,2}-\nu)\nonumber \\ && + 
\frac{5}{32}\phi(\nu_{2\,2}-\nu)+
\frac{14}{32}\phi(\nu_{3\,2}-\nu)\bigg]. 
\label{prof-combined}
\end{eqnarray}
We have verified that 
if the $F$-states are very closely spaced,
then a single profile function 
$\phi(\nu_{J_bJ_a} -\nu)$, corresponding to the $J_a \to J_b$
transition, can be used instead of $\phi_{\rm HFS}(x)$ 
\citep[see][]{ll04}.

The total source vector ${\bm S}$ in Equation~(\ref{rte-1})
is given by
\begin{equation}
{\bm S}(\tau, x, \mu)=\frac{\phi_{\rm HFS}(x){\bm S}_l(\tau,x,\mu)+
r{\bm S}_c}{\phi_{\rm HFS}(x)+r},
\label{tot-source}
\end{equation}
where the unpolarized continuum source vector ${\bm S}_c=B{\bm U}$, 
with $B$ being the Planck function and ${\bm U}=(1,0)^{T}$. The line
source vector for a two-level atom with HFS 
is given by
\begin{eqnarray}
&&{\bm S}_l(\tau,x,\mu)=\epsilon B{\bm U}+ \frac{1}{\phi_{\rm HFS}(x)}
\int^{+\infty}_{-\infty}
dx^{\prime} \nonumber \\ && \times \int^{+1}_{-1}\frac{d\mu^{\prime}}{2}
{\bf R}(x,\mu; x^{\prime},\mu^{\prime})
{\bm I}(\tau,x^{\prime},\mu^{\prime}).
\label{line-source}
\end{eqnarray}
Here $\epsilon=\Gamma_{I}/(\Gamma_{I}+\Gamma_{R})$ is the photon
destruction
probability per scattering also known as the thermalization  
parameter, with $\Gamma_I$ and $\Gamma_R$ being the 
inelastic and radiative de-excitation rates of the upper state $F_b$.
To a first approximation these rates are assumed to be the same for all the
$F$-states. 
${\bf R}(x,\mu; x^{\prime},\mu^{\prime})$ is the redistribution matrix
defined in Equation~(\ref{red_mat}) but integrated over the azimuths
 $\varphi^\prime$
of the incoming radiation. Such a simplification 
is possible due to the azimuthal symmetry of the problem. 
This redistribution matrix can be rewritten as
\begin{eqnarray}
{\bf R}_{ij}(x,{\mu};x^\prime, {\mu}^\prime)&=&\sum_{K}
{\bf{\mathcal R}}^{K}(x,x^{\prime})\nonumber \\ && \times 
{\mathcal T}^K_0(i,\mu)
{\mathcal T}^{K}_{0}(j,{\mu}^\prime).
\end{eqnarray}

The redistribution function components ${\bf{\mathcal R}}^{K}(x,x^{\prime})$
are given by
\begin{eqnarray}
&&{\bf {\mathcal R}}^{K}(x,x^\prime) = \frac{3(2J_b+1)}{2I_s+1}
\sum_{F_aF_fF_bF_{b^\prime}} (-1)^{F_f-F_a} \nonumber \\ && \times
 \cos\beta_{F_{b^\prime}F_b}  [\cos\beta_{F_{b^\prime}F_b}
(h^{\rm II}_{F_b,F_{b^\prime}})_{F_aF_f} \nonumber \\ && - 
\sin\beta_{F_{b^\prime}F_b}
(f^{\rm II}_{F_b,F_{b^\prime}})_{F_aF_f}  ] \nonumber \\ && 
\times(2F_a+1)(2F_f+1)(2F_b+1)(2F_{b^\prime}+1) \nonumber \\&& \times 
\left\lbrace 
\begin{array}{ccc}
J_a & J_b & 1\\
F_b & F_f & I_s \\
\end{array}
\right\rbrace
\left\lbrace 
\begin{array}{ccc}
J_a & J_b & 1\\
F_b & F_a & I_s \\
\end{array}
\right\rbrace \nonumber \\ && \times
\left\lbrace 
\begin{array}{ccc}
J_a & J_b & 1\\
F_{b^\prime} & F_f & I_s \\
\end{array}
\right\rbrace 
\left\lbrace 
\begin{array}{ccc}
J_a & J_b & 1\\
F_{b^\prime} & F_a & I_s \\
\end{array}
\right\rbrace \nonumber \\ && \times
\left\{
\begin{array}{ccc}
1 & 1 & K \\
F_{b^\prime} & F_b & F_a \\
\end{array}
\right\}
\left\{
\begin{array}{ccc}
1 & 1 & K \\
F_{b^\prime} & F_b & F_f \\
\end{array}
\right\}.
\label{r_mat}
\end{eqnarray}
For simplicity, we use the angle-averaged versions of the auxiliary
 functions
 $(h^{\rm II}_{F_b,F_{b^\prime}})_{F_aF_f}$ and 
$(f^{\rm II}_{F_b,F_{b^\prime}})_{F_aF_f}$.

\subsection{Decomposition of the Stokes 
vectors into the reduced basis}
Decomposition of the Stokes source vector $\bm S$ in the reduced basis
makes it independent of $\theta$.
The decomposition of $\bm S$ defined in Equation~(\ref{tot-source})
can be carried out in a way similar to the one presented 
in Section~2.1 of P2. Hence we do not repeat them here. 
The transfer equation for the reduced Stokes vector ${\bm {\mathcal I}}$ 
can be written as
\begin{equation}
\mu\frac{\partial{\bm {\mathcal I}}(\tau,x,{ \mu})}{\partial \tau}= 
(\phi_{\rm HFS}(x) + 
{r})[{\bm{\mathcal I}}(\tau, x, {\mu})-{\bm {\mathcal S}}(\tau, x)].
\label{tkq-rte}
\end{equation}
The corresponding irreducible total and 
line source vectors are given by
\begin{equation}
{\bm {\mathcal S}}(\tau,x)=\frac{\phi_{\rm HFS}(x)
{\bm{\mathcal S_{l}}(\tau,x)}+ 
{r}{\bm{\mathcal G}(\tau)}}{\phi_{\rm HFS}(x)+{r}},
\end{equation}
and
\begin{equation}
{\bm{\mathcal S_{l}}}(\tau,x)=\epsilon{\bm{\mathcal G}}(\tau) + 
\int^{+\infty}_{-\infty}
\frac{\widetilde{{\bm {\mathcal R}}}(x,x^{\prime})}{\phi_{\rm HFS}(x)}
{\bm {\mathcal J}}(\tau,x^{\prime})dx^{\prime}.
\label{source-l}
\end{equation}
Here $\widetilde{{\bm {\mathcal R}}}(x,x^{\prime})$ is a $(2\times 2)$
diagonal matrix with elements $\widetilde{{\bm {\mathcal R}}}=$diag
$({\mathcal R}^{0},{\mathcal R}^{2})$, where ${\mathcal R}^{K}$ are defined in
Equation~(\ref{r_mat}). 
 $\bm{\mathcal G}(\tau) = (B,0)^{\rm T}$ is the primary source vector,
and $\bm {\mathcal J}(\tau,x)$ is the mean intensity defined in Equation~(22)
of P2.
 
\section{The numerical methods}
\label{num-method}
Here we describe two numerical techniques to solve the reduced 
form of the transfer equation. We compare their performance
on some benchmark problems.

\subsection{Operator perturbation method}
The solution of the polarized line transfer equation defined in 
Equation~(\ref{tkq-rte})
using the polarized approximate lambda iteration (PALI) method 
 is described in Sections~3.1
and 3.2 of P2. The same equations hold good for the present
 problem also. Hence we do not  
repeat those equations here. The only difference is that 
the redistribution matrix for $J$-state interference 
is now to be replaced by the 
corresponding redistribution matrix for the $F$-state interference 
presented in this paper. Also the profile function 
is to be replaced with $\phi_{\rm HFS}(x)$.

\subsection{Scattering expansion method}
In recent years a new method based on Neumann series expansion
of the polarized source vector has been developed \citep[see][]
{hfetal09}. It is applied to a variety of astrophysical problems.
Here we describe the application of this method to the problem
at hand.

In this method, the reduced line source vector defined in 
Equation~(\ref{source-l})
is rewritten in the component form for the non-magnetic case as
\begin{eqnarray}
&&\!\!\!\!\!\!\!\!\!\!\!S^K_{0} (\tau,x) =
{G}(\tau) \delta_{K0} \delta_{00} 
+\int^{+1}_{-1} {d\mu^\prime \over 2} \nonumber \\ &&
 \!\!\!\!\!\!\!\!\!\!\!\times 
\int_{-\infty}^{+\infty} dx^\prime 
{\mathcal R^{K}(x,x^\prime)\over \phi_{\rm HFS}(x)}
\sum_{K^\prime} \Psi^{KK^\prime}_
{\rm 0}(\mu^\prime)
 I^{K^\prime}_{\rm 0} (\tau,x^\prime,\mu^\prime).\ \ \ \ \ \ \ 
\label{SEMeqn}
\end{eqnarray}
$\Psi^{KK^\prime}_{0}$ are the components of the Rayleigh
phase matrix in the reduced basis \citep[see Appendix~A of][]{hf07}.
We first consider the component $S^0_0$. Expanding the
summation over $K^\prime$ on the right-hand side of
Equation~(\ref{SEMeqn}) we obtain
\begin{eqnarray}
&&\!\!\!\!\!\!\!\!\!\!\!S^0_{0} (\tau,x) = {G}(\tau) \nonumber \\
 && \!\!\!\!\!\!\!\!\!\!\!
+\int^{+1}_{-1} {d\mu^\prime \over 2} 
\int_{-\infty}^{+\infty} dx^\prime
{\mathcal R^{0}(x,x^\prime)\over \phi_{\rm HFS}(x)}
\Psi^{00}_
{\rm 0}(\mu^\prime)
 I^{0}_{\rm 0} (\tau,x^\prime,\mu^\prime) \nonumber \\ && 
\!\!\!\!\!\!\!\!\!\!\!
+\int^{+1}_{-1} {d\mu^\prime \over 2}
\int_{-\infty}^{+\infty} dx^\prime
{\mathcal R^{2}(x,x^\prime)\over \phi_{\rm HFS}(x)}
 \Psi^{02}_
{\rm 0}(\mu^\prime)
 I^{2}_{\rm 0} (\tau,x^\prime,\mu^\prime).\ \ \ \ \ 
\label{Scalar}
\end{eqnarray}
The degree of linear polarization arising due to Rayleigh
scattering is small because of small degree of anisotropy prevailing 
in the solar atmosphere. Hence the effect of linear
polarization on Stokes $I$ can be neglected to a good
approximation. Neglecting the contribution from $I^2_0$,
in Equation~(\ref{Scalar}) we get
\begin{eqnarray}
&&\!\!\!\!\!\!\!\! \tilde S^0_{0} (\tau,x) \backsimeq
{G}(\tau) 
+\int^{+1}_{-1} {d\mu^\prime \over 2} \nonumber \\ && 
\!\!\!\!\!\!\!\! \times 
\int_{-\infty}^{+\infty} dx^\prime
{\mathcal R^{0}(x,x^\prime)\over \phi_{\rm HFS}(x)}
 \Psi^{00}_
{\rm 0}(\mu^\prime)
 I^{0}_{\rm 0} (\tau,x^\prime,\mu^\prime),
\end{eqnarray}
where $\tilde S^0_0$ denotes the approximate value of
$S^0_0$. It is the solution of a non-LTE unpolarized
radiative transfer equation and is computed using
the Frequency-by-Frequency (FBF) technique
of \citet{pl-au95}.

The polarization is computed from the higher order terms
in the series expansion. The $S^2_0$ component is given by
\begin{eqnarray}
\nonumber && \tilde S^2_{0} (\tau,x) \backsimeq
\int^{+1}_{-1} {d\mu^\prime \over 2}
\int_{-\infty}^{+\infty} dx^\prime
{\mathcal R^{2}(x,x^\prime)\over \phi_{\rm HFS}(x)}\\&&
\nonumber\times \Psi^{20}_
{\rm 0}(\mu^\prime)
 \tilde I^{0}_{\rm 0} (\tau,x^\prime,\mu^\prime)\\&&
\nonumber+\int^{+1}_{-1} {d\mu^\prime \over 2}
\int_{-\infty}^{+\infty} dx^\prime
{\mathcal R^{2}(x,x^\prime)\over \phi_{\rm HFS}(x)}\\&&
\times \Psi^{22}_
{\rm 0}(\mu^\prime)
 \tilde I^{2}_{\rm 0} (\tau,x^\prime,\mu^\prime).
\label{s2comp}
\end{eqnarray}
Retaining only the contribution from $\tilde I^0_0$
on the right-hand side of Equation~(\ref{s2comp}) we obtain the
single scattering approximation
to the polarized component of the source vector as
\begin{eqnarray}
\nonumber && [\tilde S^2_{0} (\tau,x)]^{(1)} \backsimeq
\int^{+1}_{-1} {d\mu^\prime \over 2}
\int_{-\infty}^{+\infty} dx^\prime
{\mathcal R^{2}(x,x^\prime)\over \phi_{\rm HFS}(x)}\\&&
\times \Psi^{20}_
{\rm 0}(\mu^\prime)
 \tilde I^{0}_{\rm 0} (\tau,x^\prime,\mu^\prime). 
\end{eqnarray}
The superscript (1) denotes single (first) scattering. This
solution serves as a starting point for the computations of higher
order scattering terms. Thus the iterative sequence of SEM can be
represented by
\begin{eqnarray}
&& [\tilde S^2_0(\tau,x)]^{(n)} \backsimeq
[\tilde S^2_{0} (\tau,x)]^{(1)}\nonumber \\ &&
+\int^{+1}_{-1} {d\mu^\prime \over 2}
\int_{-\infty}^{+\infty} dx^\prime
{\mathcal R^{2}(x,x^\prime)\over \phi_{\rm HFS}(x)}\nonumber \\ &&
\times \Psi^{22}_
{\rm 0}(\mu^\prime)
 [\tilde I^{2}_{\rm 0} (\tau,x^\prime,\mu^\prime)]^{(n-1)},
\end{eqnarray}
where the superscript $(n)$ denotes the $n^{\rm th}$ scattering.
The iterative cycle is continued until the required
convergence criteria is met.

\begin{figure}
\centering
\includegraphics[width=7.0cm,height=7.0cm]{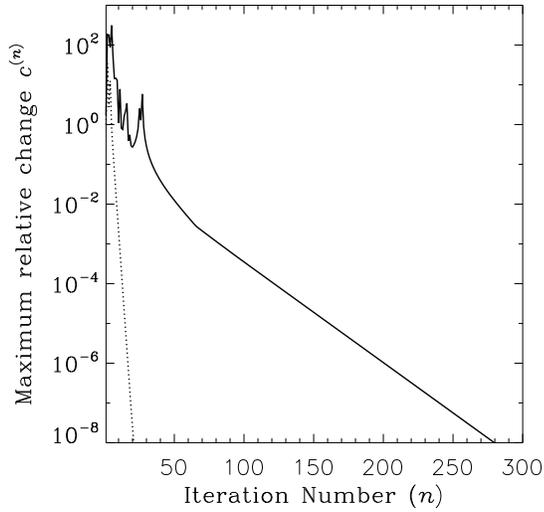}
\caption{\footnotesize Comparison of PALI (solid line) and
scattering expansion method (dotted line).
The model parameters are given in the text. A convergence
criteria of $10^{-8}$ is used.}
\label{mrcplot}
\end{figure}

In the following we compare the performance of these two
numerical methods by plotting the maximum relative correction
defined as
\begin{equation}
c^{(n)}={\rm max}\{c_1^{(n)},c_2^{(n)}\} < 10^{-8},
\end{equation}
where
\begin{equation}
c_1^{(n)}= {\rm max}_{\ \tau,x,\mu}\bigg\{\frac{|\delta S_I^{(n)}
(\tau,x,\mu)|}{|\bar{S}^{(n)}_I(\tau,x,\mu)|}\bigg\},
\end{equation}
and
\begin{equation}
c_2^{(n)}= {\rm max}_{\ x,\mu}\bigg\{\frac{P^{(n)}(x,\mu)-P^{(n-1)}
(x,\mu)}{P^{(n-1)}(x,\mu)}\bigg\},
\end{equation}
as a function of the iteration number as shown in Figure~\ref{mrcplot}.
In the above equations $P=[Q/I]$ is the degree of
linear polarization and
$\bar{S}_I^{(n)}=\frac{1}{2}[S^{(n)}_{I}+S^{(n-1)}_{I}]$.

Figure~\ref{mrcplot} is computed for a test problem defined by the
model parameters $(T,a,\epsilon,r,B)=(2\times10^{10},2\times10^{-3},
10^{-4},0,1)$ where $T$ is the optical thickness of the self
emitting slab and $a$ is the damping parameter of the upper level $J_b$. 
From the figure one can clearly see that the convergence rate of
the SEM is larger by several factors compared to the PALI method.
{ The reason for the PALI method being slow is that the source
function corrections are computed iteratively from an
approximate initial guess and then the approximate lambda
operator is perturbed until the source function corrections fall below a
convergence criteria}. On the other hand, the initial guess in the
SEM for polarized line formation is the single scattered solution
itself (which already contains the physical characteristics of
the scattering mechanism under consideration). For this reason
SEM takes just a few iterations to converge to the same level of
accuracy as the PALI method. Further, SEM is easy to implement
for problems of any physical and/or numerical complexity. This
makes the SEM a method of choice. For a detailed comparison
of PALI and SEM we refer to \citet[][]{sam11} and \citet[][]{sup12}.
The simple Lambda iteration for polarization and the SEM are 
essentially similar. In the lambda iteration, a source vector correction is 
computed at each iteration, and the current source vector is 
updated until convergence is reached. 
In the SEM, each iteration can be seen as contributing 
a higher order scattering term to the series expansion of polarized component
of the source vector. This  component  
is updated by adding successively higher order terms 
in the scattering expansion of the source vector. These points are clearly 
explained respectively in \citet[][see the discussion 
following their Equation~(28)]{jtbms99}, and 
\citet[][see the discussion following their Equation~(36)]{hfetal09}.
\section{Results and discussion}
\label{results}
In this section we present the results computed for a standard two-level
atom model with $F$-state interference using the PRD matrix 
presented in this paper. Isothermal constant property media
characterized by $(T,a,\epsilon,r,B)$ are used. The slabs are assumed 
to be self-emitting. 

The results are presented for transitions centered at hypothetical  
wavelengths arising due to HFS of the 
$J_b=3/2$ and $J_a=1/2$ levels of a two-level atom with nuclear spin $I_s=3/2$.
Due to the hyperfine interactions the upper $J$-state splits into four $F$-states
 with $F_b=0,1,2,3$, and the lower $J$-state splits into
$F_a=1,2$. Owing to the selection rule $\Delta F=0,\pm1$, these $F$-states 
produce six radiative transitions involving them (see Table~{\ref{table-1}}). 
For simplicity 
the Doppler width of all the lines 
is taken to be constant at $\Delta\lambda_D=25$ m\AA. 
In the transfer computations, a grid resolution of 
$(N_d,N_x,N_\mu)$ = (5, 417, 5) is generally used, where $N_d$ 
is the number of depth points per decade in the logarithmically spaced 
$\tau-$grid. The first depth point is taken as $\tau_{\rm min}=10^{-2}$. 
$N_x$ is the total number of frequency points covering the full line profile.
$N_\mu$ is the number of co-latitudes $\theta(\mu)$, taken as the 5 
points of a Gauss-Legendre quadrature. 

\subsection{$F$-state interference effects in the case of single scattering}
\begin{figure}
\centering
\includegraphics[width=8.0cm,height=7.0cm]{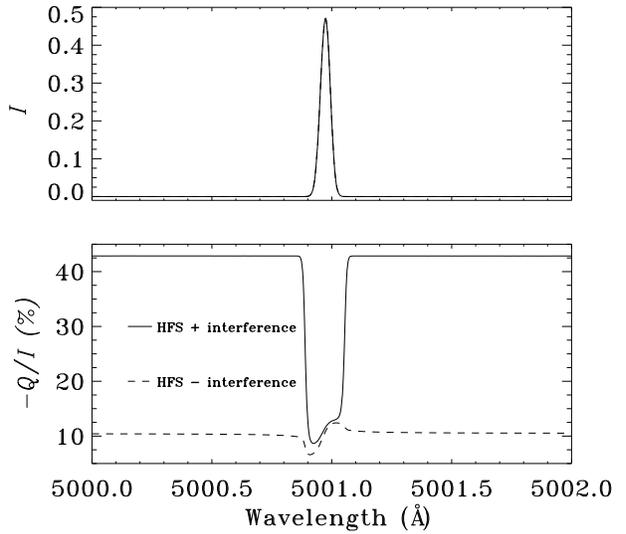}
\caption{\footnotesize The profiles of the intensity $I$ and the fractional
polarization $Q/I$, plotted for a hypothetical line system with hyperfine
structure splitting. Solid line represents the $Q/I$ with $F$- state 
interference and dashed line represents $Q/I$ without $F$-state interference.
Single 90\textdegree\  scattering is assumed at the extreme 
limb ($\mu=0$). The model parameters are $a=0.002$, 
the Doppler width $\Delta\lambda_{\rm D}=0.025$\ \AA.}
\label{sing-scat}
\end{figure}

In this section we study the behavior of the 
$F$-state interference PRD matrix derived in Section~\ref{redmatrix}
by computing the scattered profiles in a single scattering event. 
The results in Figure~{\ref{sing-scat}}
are computed for a 90$^{\circ}$ single scattering event. 
 This is done by giving as input an unpolarized
beam of light incident on the scattering atom at 
$\mu^\prime=-1$ and observing the scattered ray at $\mu=0$
in the scattering plane (see P1 for details on computing
polarization profiles in a 90$^\circ$ single scattered event).
The dashed line in Figure~{\ref{sing-scat}} is computed
by ignoring the interference effects, whereas the solid line
is computed by taking account of the interference effects
between the $F$-states. 
Profile similar to the solid line can also be seen in \citet[][]{fluri03}
and \citet[][]{holz05} where plots of the wavelength 
dependent polarizability factor $W_{2}(\lambda)$ are shown. 
In the single scattering case, the profiles of the $W_2(\lambda)$ and 
the $Q(\lambda)/I(\lambda)$
are similar in shape but differ only in magnitude (see below).

\subsubsection{Principle of spectroscopic stability 
for $F$-state interference}
\label{pss}
It is well known that the principle of spectroscopic stability 
provides a useful tool to check any theory of quantum interference.
This was first discussed in the context of scattering
 polarization and applied in detail in \citet[][]{jos94}
\citep[see also][]{s97,ll04}. In this paper, we  
apply it to the case of $F$-state interference
arising due to the nuclear spin $I_s$. According to the principle of
spectroscopic stability, 
in the limit of vanishing HFS in a two-level atom,
the theory of $F-$state interference should reduce to the 
standard two-level atom theory without HFS.
This can be verified by computing the polarizability factor $W_2$
and in turn the fractional polarization $Q/I$
in the limit of vanishing $F$-states. 
The value of $W_2$ in this asymptotic limit (which can be obtained
 by neglecting the 
$I_s$) can be computed as described in \citet{s97} with 
\begin{eqnarray}
 (W_2)_{\rm asym} = \frac{\left\{
\begin{array}{ccc}
1 & 1 & 2 \\
J_{b} & J_b & J_a \\
\end{array}
\right\} \left\{
\begin{array}{ccc}
1 & 1 & 2 \\
J_{b} & J_b & J_f \\
\end{array}
\right\}}{\left\{
\begin{array}{ccc}
1 & 1 & 0 \\
J_{b} & J_b & J_a \\
\end{array}
\right\}\left\{
\begin{array}{ccc}
1 & 1 & 0 \\
J_{b} & J_b & J_f \\
\end{array}
\right\}}.
\end{eqnarray}
For the particular case of $J_a=1/2 \to J_b=3/2 \to J_f=1/2$
scattering transition,
$(W_2)_{\rm asym} =0.5$. Hence  $W_2(\lambda)$ 
is expected to approach 0.5 in the very far wings 
\citep[see Figure~2 of][]{s97}. In the 
$90^{\circ}$ single scattering case, the $Q/I$
and the $W_2(\lambda)$ are related through the formula
\citep[see][]{ll04}
\begin{equation}
Q(\lambda)/I(\lambda)=\frac{3W_2(\lambda)}{4-W_2(\lambda)}.
\label{w2-qbyi}
\end{equation}
The above formula gives in the far wings a value of $Q/I=0.428$
for $(W_2)_{\rm asym}=0.5$. 

From Figure~{\ref{sing-scat}}, we can see that
the solid curve reaches an asymptotic value of $42.8\%$ as demanded 
by the principle of spectroscopic stability, whereas the dashed line
reaches about $10\%$ in the far wings, thereby violating the principle of 
spectroscopic stability. These arguments show that in the formulation 
of the redistribution matrix, the inclusion of interference effects
between the $F$-states is essential.

\subsection{Effects of $F$-state interference in multiply
scattered Stokes profiles}
In this section we present the results obtained by solving 
the transfer equation 
including the $F$-state interference. In the particular case of 
optically thin slabs, it can be shown, by choosing appropriate 
geometric 
arrangement, that the multiply scattered solution approaches 
single scattered solution thus proving that we have correctly
incorporated the $F$-state redistribution matrix 
in the line transfer code.  See P2 for more details 
regarding single scattering in a thin atmospheric slab. 

\begin{figure}
\centering
\includegraphics[width=8.0cm,height=13.0cm]{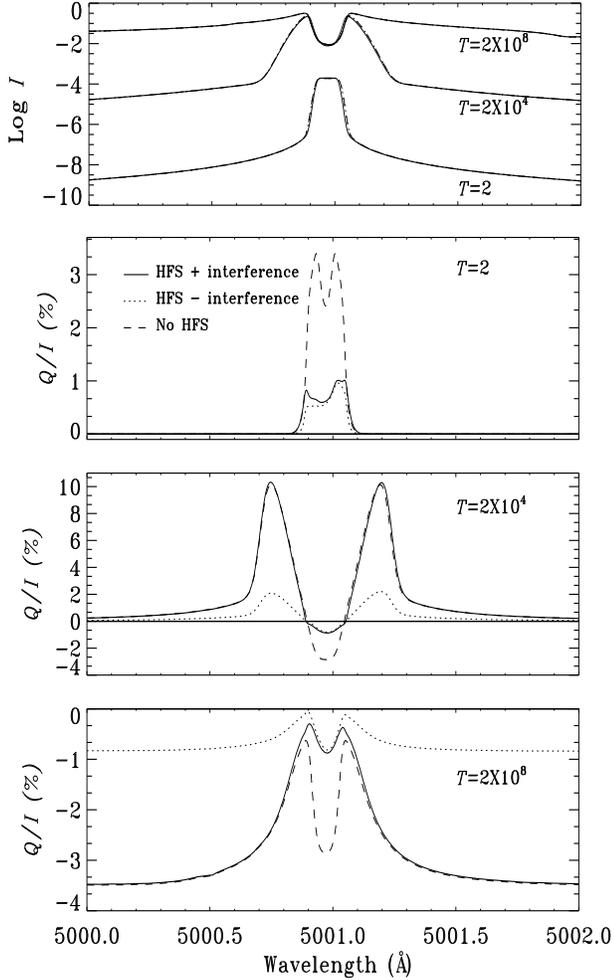}
\caption{\footnotesize Comparison between the multiply scattered
emergent Stokes profiles computed for different atomic systems
as indicated in panel 2. The model parameters are 
$(a,\epsilon,{r},B)=(2\times10^{-3}, 10^{-4}, 0, 1)$.
The line of sight is given by $\mu=0.047$. The wavelength
positions of the six components are given in Table~1. The spacing
between the hyperfine structure components is taken to be the same as 
those corresponding to the Na\,{\sc i} D$_2$ line.}
\label{multiply-scat}
\end{figure}

When the optical thickness of the medium is large, multiple
scattering effects come into play. Figure~{\ref{multiply-scat}} shows 
one such example, where the emergent Stokes profiles are computed 
for different optical thicknesses. The dashed line in this figure is
computed by neglecting HFS. This is the standard two-level atom case
which results in a single radiative transition. The dotted line is 
computed
with HFS but without interference between the $F$-states.
In this case the six radiative transitions arising due to HFS
are treated independently. 
The solid line is computed taking account of the $F$-state 
interference. This comprises of  six interfering radiative
transitions between the $F$-states. 
The three line types in this figure are
quite similar to each other in shape but differ prominently in amplitude.

For $T=2$, the atmospheric slab is effectively thin and the $Q/I$
profiles for both solid and dotted lines
have a structuring within the line core which is different from that of 
the dashed line. This is due to the HFS of the given $J$-level.
As the optical thickness increases, such a structuring gets 
smoothened out and the shape (not the amplitude) 
of the solid and dotted line profiles
resemble more closely with the dashed line profiles. 
 
In the case of effectively thick atmospheric slabs ($T>2$),
two peaks are seen on either side of the line center
arising due to PRD effects and are known as PRD wing peaks.
In the line core, the solid and dotted lines nearly coincide
whereas the dashed line differs from these two. This shows that
the depolarization in the line core is purely due to HFS, 
irrespective of the interference effects between the $F$-states being 
included. 
In the wings, the solid line and the dashed line coincide whereas
the dotted line differs significantly. Upon comparing the solid and 
dotted lines,
it is evident that the interference effects show up in the line wing 
PRD peaks
like in the case of $J$-state interference.
However the $J$-state interference effects are seen
even beyond the PRD wing peaks unlike the case of $F$-state interference.
When $F$-state interference is taken into account the $Q/I$ 
in the wings reaches the value of the single line case
as expected from the principle of spectroscopic stability 
(see Section~{\ref{pss}}).
But when interference is neglected, the dotted and dashed lines differ
considerably in the wings which can be seen as a violation of the 
principle of spectroscopic stability. Thus 
the principle of spectroscopic stability serves as a powerful tool
to check the correctness of our formulation not only in the
case of single scattering but also in the
radiative transfer computations.

Though such significant signatures of HFS and $F$-state interference
are seen in $Q/I$, the intensity $I$ remains
unaffected by these effects. 

\subsection{Comparison with wavelength dependent polarizability 
theory of Stenflo}
\begin{figure}
\centering
\includegraphics[width=8.0cm,height=7.0cm]{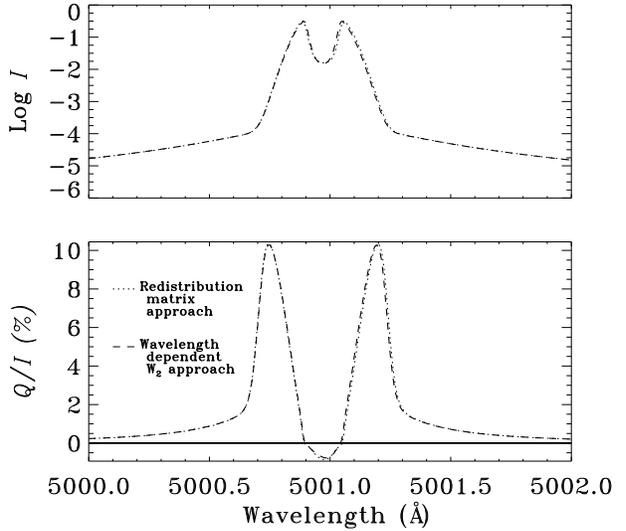}
\caption{\footnotesize Comparison between the redistribution matrix 
theory (dotted line) and wavelength dependent polarizability
factor $W_2 (\lambda)$ theory of Stenflo (dashed line).
The optical thickness of the atmospheric slab is $T=2\times 10^4$.
The other model parameters are the same as in Figure~{\ref{multiply-scat}}.}
\label{w2-rt}
\end{figure}

In  this section we compare our redistribution matrix approach 
and the wavelength dependent polarizability $W_2(\lambda)$
theory for the case of $F$-state interference 
presented in \citet[][]{s97} and used in \citet{fluri03}
 and \citet{holz05}.
The comparison is shown in Figure~{\ref{w2-rt}}. 
The dotted lines show the profiles computed using the
exact PRD $F$-state interference theory presented in 
Section~{\ref{theory}}. 
This is our redistribution matrix approach.
The dashed lines show the profiles computed using the 
$W_2(\lambda)$ approach.
The values of the $W_2(\lambda)$ are calculated from
 Equation~(\ref{w2-qbyi})
using the $(Q/I)$ plotted in Figure~{\ref{sing-scat}} 
(solid line). 
To use the $W_2(\lambda)$ in radiative transfer computations
 we replace
the redistribution matrix $\mathcal{R}^{K}(x,x^{\prime})$ in
 Equation~(\ref{r_mat})
by
\begin{eqnarray}
W_{K}(\lambda)[R^{\rm II -A}(3/2 \to 1/2)],
\end{eqnarray}
where $R^{\rm II -A}(J_b \to J_f)$ is
the angle-averaged frequency 
redistribution function of \citet{hum62} for a line 
centered at $\lambda_{J_bJ_f}$ corresponding 
to the $J_b \to J_f$ transition. 
For the hypothetical case under study, we have assumed the 
$F$-states to be very closely spaced. Under such an assumption,
a single redistribution function computed for the $J=3/2 \to 1/2$
transition can be used to represent all the $F$-state transitions.
However if the $F$-states are far apart then the redistribution 
function 
needs to be computed for each of the $F_b \to F_f$ transition.
In such a case, the redistribution matrix $\mathcal{R}^{K}(x,x^{\prime})$
takes the following form in the $W_2(\lambda)$ approach:
\begin{eqnarray}
 W_{K}(\lambda)\sum_{F_bF_f}[R^{\rm II -A}(F_b \to F_f)].
\end{eqnarray}
The polarizability factor
$W_{0}(\lambda)=1$, and $W_{2}(\lambda)$ is the
wavelength-dependent $W_2$ factor calculated from 
Equation~(\ref{w2-qbyi}). For the closely spaced 
$F$-states a common absorption profile function $\phi(x)$ 
corresponding to the $J_a=1/2 \to J_b=3/2$
transition is used. But in the case of widely spaced $F$-states, the 
$\phi(x)$ has to be taken as the sum of all the individual $F_a \to F_b$ 
absorption profile functions.
As seen from Figure~{\ref{w2-rt}} both the redistribution matrix approach and the
$W_2(\lambda)$ approach give nearly the same results.
\section{Conclusions}
\label{conclusions}
In the present paper we have extended the $J$-state interference 
formulation discussed in P1 and P2 to the case of $F$-state interference.
The treatment is restricted to the collisionless and non-magnetic regime.
The decomposition technique presented in \citet{hf07} is applied 
to the $F$-state interference problem. It helps to incorporate
the $F$-state interference redistribution matrix into the 
reduced form of the line radiative transfer equation.
The transfer equation is solved using the traditional PALI
and the scattering expansion method by suitably adapting them
to handle the $F$-state interference problem. The SEM is found 
to be more efficient and faster than the PALI method.

The Stokes profiles computed by taking account of HFS
are similar to the profiles of a single line arising 
from a two-level atom model without HFS. The HFS causes a 
depolarization of $Q/I$ in the line core irrespective
of whether the $F$-state interference is taken into account 
or not. Like the $J$-state interference, the $F$-state 
interference affects mainly the line wing PRD peaks. 
We also show that when interference effects are neglected,
the principle of spectroscopic stability is violated
in both single scattered and multiple scattered profiles.
Using the fractional polarization $Q/I$ in the 
$90^\circ$ single scattering case, we can numerically estimate the 
wavelength dependent
polarizability factor $W_2(\lambda)$. The $W_2(\lambda)$ 
so computed can then be used in the 
transfer equation to compare with our exact redistribution matrix 
approach. The two approaches are found to give identical
emergent Stokes profiles.

\begin{table*}
\caption{ Wavelengths (\AA) of $F$-state transitions for
a hypothetical atomic system}\ \\
\label{table-1}
\centering
\begin{tabular}{|c|cccc|}
\hline
\ & $F_b=0$ & $F_b$=1 & $F_b$=2 & $F_b$=3\\
\hline 
$F_a=1$ & 5000.96093 & 5000.96075 &  5000.96036 & N.A \\
$F_a=2$ & N.A & 5000.98125 & 5000.98086 &  5000.98018 \\
\hline
\end{tabular}
\end{table*}

\end{document}